\documentclass[aps,pra,showpacs,amssymb,twocolumn]{revtex4-1}
\usepackage{graphicx}
\usepackage{appendix} 
\usepackage{hyperref}
\usepackage{bm,amsmath}
\usepackage{dcolumn}

\begin{document}

\vspace{0.2in}

\title{Efficient error models for fault-tolerant architectures and the Pauli twirling approximation}

\author{Michael R. Geller and Zhongyuan Zhou}
\affiliation{Department of Physics and Astronomy, University of Georgia, Athens, Georgia 30602, USA}

\date{\today}

\begin{abstract}
The design and optimization of realistic architectures for fault-tolerant quantum computation requires error models that are both reliable and amenable to large-scale classical simulation. Perhaps the simplest and most practical general-purpose method for constructing such an error model is to twirl a given completely positive channel over the Pauli basis, a procedure we refer to as the Pauli twirling approximation (PTA). In this work we test the accuracy of the PTA for a small stabilizer measurement circuit relevant to fault-tolerant quantum computation, in the presence of both intrinsic gate errors and decoherence, and find excellent agreement over a wide range of physical error rates. The combined simplicity and accuracy of the PTA, along with its direct connection to the $\chi$ matrix of process tomography, suggests that it be used as a standard reference point for more refined error model constructions.
\end{abstract}


\maketitle

\section{Pauli twirling approximation}
\label{introduction section}

The principal obstacle to large-scale quantum computation is the introduction errors caused by decoherence, noise, leakage to non-computational states, incorrect implementation of quantum gates, 
qubit loss, and inaccurate state initialization and measurement. The standard approach for mitigating these errros is to use a fault-tolerant error-correction protocol, which enables arbitrarily large computations as long as the strength of the errors are below a threshold value \cite{AharonovProcIEEE97,KitaevRussMathSurv97,KnillProcRoySocLondA98,AharonovSIAMJC08,FowlerPRL12} and are not overly correlated in space or time \cite{TerhalPRA05,AliferisQIC06,AharonovPRL06,NovaisPRA08,NovaisPRL13}. The fault-tolerant error threshold is a measure of the robustness of a quantum computing platform and an estimate of its value is one of the most important tasks for practical quantum computer design. 

A straightforward approach for calculating logical error rates and associated error thresholds would be to do a full Hilbert space simulation of quantum codes of increasing size, in the presence of decoherence and other errors, but this approach quickly becomes intractable. This difficulty can be circumvented by using special error models that are efficient to simulate classically. The existence of a broad class of these {\it efficient} error models, which includes the Pauli and Clifford channels, is provided by the Gottesman-Knill theorem: This theorem states that any quantum circuit consisting only of Clifford-group unitaries and measurement in the Pauli basis can be simulated classically in polynomial time \cite{GottesmanArxiv9705052,AaronsonPRA04}. The circuits used to implement stabilizer-based error detection and correction---in the absence of any errors---are important examples. However, after including decoherence and intrinsic gate errors, which are required to assess fault-tolerance and calculate error thresholds, the simulations are no longer efficient. By {\it intrinsic} we mean an error, such as a  unitary qubit rotation by the wrong angle, which does not result from noise or decoherence. The resulting inefficiency of classical simulation is the main motivation for the widely used stochastic approach of modeling nonideal stabilizer circuits by a sequence of one- and two-qubit operations, each of which consists the intended ideal Clifford gate, possibly followed by a unitary ``error" randomly drawn from the Pauli or Clifford basis according to some probability distribution. This approach raises several important questions for the design and optimization of realistic quantum computing architectures: What set of error operations---which can also include Pauli-basis measurement---should be allowed? How should the probability distribution over that set be determined? And how reliable is this entire approach?

A first step towards answering these questions was taken recently by Magesan {\it et al.}~\cite{MagesanPRA13} and by Guti\'{e}rrez {\it et al.}~\cite{GutierrezPRA13}. The focus of Refs.~[\onlinecite{MagesanPRA13}] and [\onlinecite{GutierrezPRA13}] is similar: Both find the Pauli or Clifford channel closest (according to a chosen norm) to a given target error channel subject to a constraint that the approximation upper-bounds some measure of the error. Here we take a different approach by attempting to directly assess the accuracy of one of the simplest known error model constructions---what can be called the Pauli twirling approximation---when applied to a realistic stabilizer measurement circuit.

The completely positive time evolution of an $N \! \times \! N$ density matrix can be represented by
\begin{equation}
\rho \rightarrow \Lambda\big(\rho\big) = \sum_{m} E_m \, \rho \, E_m^\dagger,
\label{CP map}
\end{equation}
where the $E_m$ are $N \! \times \! N$ Kraus matrices. The process $\Lambda$ does not need to be trace preserving. {\it Twirling} \cite{BennettPRA96,DurPRA05,EmersonSci07,DankertPRA09,LopezPRA10} a process $\Lambda$ refers to the pre-multiplication of the input state by an operation $A$, running the original process $\Lambda$, and then post-multiplying by $A^{-1}$; this procedure is then averaged over a set of $K$ operations ${\cal A} = \lbrace A_k \rbrace_{k=1}^K.$ Denoting the twirled process by ${\tilde \Lambda}$, we have 
\begin{equation}
{\tilde \Lambda}\big(\rho\big) \equiv \frac{1}{K} \sum_{A \in {\cal A} } A^{-1} \, \Lambda\big(A \rho A^{-1} \big) \, A .
\label{twirling definition}
\end{equation}

In this work we will be interested in the case where $\rho$ is an $n$-qubit density matrix, $N=2^n$, $\Lambda$ characterizes the residual {\it error} associated with an ideal operation $O$, and ${\cal A}$ is the $n$-qubit Pauli basis
\begin{equation}
{\cal A}_n \equiv \lbrace I, X , Y , Z  \rbrace^{\otimes n}
\label{pauli basis definition}
\end{equation}
consisting of the $4^n$ distinct tensor products of Pauli matrices $I$, $X$, $Y$, and $Z$. Twirling (\ref{CP map}) over the Pauli basis (\ref{pauli basis definition}) results in a new map ${\tilde \Lambda}$ that is diagonal in the Pauli basis \cite{BennettPRA96,DurPRA05,EmersonSci07,DankertPRA09,LopezPRA10},
\begin{equation}
{\tilde \Lambda}\big(\rho\big) = \sum_{A \in {\cal A}_n}  p_{\scriptscriptstyle A} \, A \, \rho \, A^\dagger,
\label{diagonal map}
\end{equation}
where the $p_{\scriptscriptstyle A}$ are uniquely determined from the Kraus matrices $E_m$. If the process (\ref{CP map}) is trace preserving, then 
\begin{equation}
\sum_{A \in {\cal A}_n}  p_{\scriptscriptstyle A} = 1.
\label{probability conservation condition}
\end{equation}
However there are important error processes, such as those including leakage, where $\sum_{A}  p_{\scriptscriptstyle A} < 1.$ The expression (\ref{diagonal map}) shows that the PTA maps every process to an $n$-qubit Pauli channel. In this paper $A$ refers to an element of the Pauli basis ${\cal A}_n$, with $n \ge 1$, whereas $a$ always represents an element of ${\cal A}_1$:
\begin{equation}
a \in  \lbrace I, X , Y , Z  \rbrace.
\label{a definition}
\end{equation}
Thus we can write every $A \in {\cal A}_n$ as
\begin{equation}
A = a_1 \otimes a_2 \otimes \cdots \otimes a_n, \ \  {\rm where} \ \  a_i \in \lbrace I, X , Y , Z  \rbrace.
\end{equation}
Twirling over the larger Clifford group has also been considered for the purpose of simplifying experimental process tomography \cite{DurPRA05,EmersonSci07}.

There are several features that make the PTA especially practical for efficient error model construction:
\begin{enumerate}

\item The approxmation is generally applicable to any single- or multi-qubit error process. It can be used to model both decoherence and intrinsic gate errors. Examples of each are given in Sec.~\ref{examples section}.

\item It is straightforward to apply, and, in many cases, leads to simple analytic formulas for the Pauli-error probabilities $p_{\scriptscriptstyle A}$.

\item The resulting twirled channel is itself very simple, and contains far fewer terms than the most general classically efficient channel allowed by the Gottesman-Knill theorem. This simplifies the large-scale Monte Carlo simulation of error-corrected quantum computation with codes such as Autotune \cite{FowlerPRX12}.

\item The probabilities $p_{\scriptscriptstyle A}$ can also be directly measured experimentally without complete process tomography: They are the diagonal elements of the $\chi$ matrix in the Pauli basis \cite{DurPRA05,EmersonSci07,DankertPRA09,LopezPRA10}.
\end{enumerate}

An important ingredient of the error model constructions of Refs.~[\onlinecite{MagesanPRA13}] and [\onlinecite{GutierrezPRA13}] is the bounding property, that the approximations do not overestimate (some measure of) the channel fidelity. However, upper-bounding the channel error does not necessarily upper-bound the {\it logical} error probability of a fault-tolerant computation, and in this work we are more concerned with the reliability of simulated logical error rates, especially well below the error threshold. Although the PTA {\it can} underestimate the logical error  rate, we find that its performance when applied to the four-qubit stabilizer circuit of Sec.~\ref{Bell state section} is already suffifciently impressive that there is little room for further improvement---at least for the small system considered here. We will return to this issue in Sec.~\ref{conclusion section}.

The organization of this paper is as follows: In Sec.~\ref{examples section} we provide two example applications of the PTA. The first is to a somewhat general model of decoherence that includes non-Markovian dephasing, extending the results of Refs.~[\onlinecite{SarvepalliProcRoySocA09}] and [\onlinecite{GhoshFowler&GellerPRA12}]. The second is to a high-fidelity controlled-Z ({\sf CZ}) gate designed for superconducting qubits. In Sec.~\ref{Bell state section} we consider a primitive error-correction protocol, that of preserving a pair of qubits in a single given Bell state. We calculate the error-correction failure probability in the presence of decoherence and intrinsic gate errors, both exactly and with the PTA, and find remarkable agreement order a wide range of physical error rates. Sec.~\ref{conclusion section} contains our conclusions.

\section{PTA Examples}
\label{examples section}

In this section we apply the PTA to typical one-qubit (Sec.~\ref{decoherence section}) and two-qubit (Sec.~\ref{CZ gate section}) error channels.

\subsection{Qubit decoherence}
\label{decoherence section}

Here we apply the PTA to a model of single-qubit decoherence defined by the map
\begin{eqnarray}
\rho &=& 
\begin{pmatrix}
1-\rho_{11} & \rho_{01} \nonumber \\
\rho_{01}^* &  \rho_{11} 
\end{pmatrix} \\
&\rightarrow &
\begin{pmatrix}
1-\rho_{11} e^{-t/T_1} & \rho_{01} e^{-t/2T_1} e^{-(t/T_\phi)^{1+\alpha}} \\
\rho_{01}^* e^{-t/2T_1} e^{-(t/T_\phi)^{1+\alpha}} &  \rho_{11} e^{-t/T_1} 
\end{pmatrix} \! , \ \ \ \ \
\label{qubit decoherence model}
\end{eqnarray}
which includes pure dephasing by classical noise with a power spectrum 
\begin{equation}
S(f) = {\rm const} \times \frac{1}{f^\alpha},
\end{equation}
and which is non-Markovian (the time-evolutions do not form a semigroup) if $\alpha \neq 0$. This is described by the channel
\begin{equation}
\Lambda\big(\rho\big) = \sum_{m=1}^3 E_m \, \rho \, E_m^\dagger,
\label{decoherence channel}
\end{equation}
with Kraus matrices
\begin{eqnarray}
E_1 &=& \begin{pmatrix} 1 & 0 \\ 0 &  \sqrt{1-\gamma-\lambda} \end{pmatrix} \nonumber \\
&=& \frac{1 + \sqrt{1-\gamma-\lambda}}{2} \, I 
 +   \frac{1 - \sqrt{1-\gamma-\lambda}}{2} \, Z , \nonumber \\
E_2 &=& \begin{pmatrix} 0 & \sqrt{\gamma} \\ 0 &  0 \end{pmatrix} 
= \frac{\sqrt{\gamma}}{2} \, X +  \frac{i\sqrt{\gamma}}{2} \, Y ,\nonumber \\
E_3 &=& \begin{pmatrix} 0 & 0 \\ 0 &  \sqrt{\lambda}  \end{pmatrix} 
= \frac{\sqrt{\lambda}}{2} \, I - \frac{\sqrt{\lambda}}{2} \, Z ,
\label{Kraus matrices for decoherence}
\end{eqnarray}
where
\begin{equation}
\gamma \equiv 1 - e^{-t_{\rm step} /T_1} 
\end{equation}
and
\begin{equation}
\lambda \equiv e^{-t_{\rm step}/T_1} \bigg[1  - e^{-2(t_{\rm step} /T_\phi)^{1+\alpha} } \bigg]. 
\end{equation}
Here we have assumed that (\ref{qubit decoherence model}) is to be applied to an individual step in a quantum circuit, with time duration $t_{\rm step}$. The decoherence model (\ref{qubit decoherence model}) reduces to standard $T_{1,2}$ decoherence in the limit $\alpha \rightarrow 0$ and
\begin{equation}
\frac{1}{T_\phi}  \rightarrow  \frac{1}{T_2} - \frac{1}{2T_1}. 
\end{equation}

To apply the PTA to (\ref{qubit decoherence model}), we expand the Kraus matrices (\ref{Kraus matrices for decoherence}) in the single-qubit Pauli basis [see (\ref{a definition})]
\begin{equation}
E_m = \! \! \! \sum_{a \in \lbrace I, X , Y , Z  \rbrace} \Gamma_{ma} \, a ,
\label{pauli basis kraus expansion}
\end{equation} 
where the coefficients $\Gamma_{ma}$ are given in Table \ref{decoherence gamma table}. Then (\ref{decoherence channel}) is equivalent to
\begin{equation}
\Lambda\big(\rho\big) = \sum_{a,a' \in {\cal A}_1} 
\bigg( \sum_m \Gamma_{m a} \, \Gamma_{m a'}^* \bigg) \, a \, \rho \, a'^\dagger.
\label{decoherence channel in pauli basis}
\end{equation}
The PTA replaces (\ref{decoherence channel in pauli basis}) by the diagonal Pauli channel
\begin{equation}
{\tilde \Lambda}\big(\rho\big) = \! \! \! \sum_{a  \in \lbrace I,X,Y,Z \rbrace } 
p_a \, a \, \rho \, a^\dagger ,
\label{single-qubit channel in pauli basis after PTA}
\end{equation}
with
\begin{equation}
p_a \equiv \sum_m \big| \Gamma_{m a} \big|^2.
\label{single-qubit channel probabilities}
\end{equation}
Therefore, in the PTA the single-qubit decoherence model (\ref{qubit decoherence model}) becomes
\begin{eqnarray}
\rho  &\rightarrow&  \big(1-p_{\scriptscriptstyle X} - p_{\scriptscriptstyle Y} - p_{\scriptscriptstyle Z}  \big) \, \rho  \nonumber \\
&+& p_{\scriptscriptstyle X} \, X \rho \, X 
+ p_{\scriptscriptstyle Y} \, Y \rho \, Y + p_{\scriptscriptstyle Z} \, Z \rho \, Z , 
\label{conserving single-qubit pauil channel}
\end{eqnarray}
where
\begin{eqnarray}
p_{\scriptscriptstyle X} &=& \frac{\gamma}{4} = \frac{1 - e^{-t_{\rm step}/T_1}}{4} \approx \frac{t_{\rm step}}{4 T_1} , \nonumber \\
p_{\scriptscriptstyle Y} &=& \frac{\gamma}{4}  = \frac{1 - e^{-t_{\rm step}/T_1}}{4} \approx \frac{t_{\rm step}}{4 T_1} , \nonumber \\
p_{\scriptscriptstyle Z} &=& \frac{1}{2} - \frac{\gamma}{4} - \frac{ \sqrt{ 1 - \gamma - \lambda} }{2} \nonumber \\
&=& \frac{1}{2} - \frac{1 - e^{-t_{\rm step}/T_1}}{4} - \frac{e^{-t_{\rm step}/2T_1} e^{-(t_{\rm step}/T_\phi)^{1+\alpha}} }{2} \nonumber \\
&\approx& \frac{1}{2} \bigg( \frac{t_{\rm step}}{T_\phi} \bigg)^{1+\alpha} \! \! \!.
\label{PTA decoherence probabilities}
\end{eqnarray}
The approximate expressions in (\ref{PTA decoherence probabilities}) apply when $t_{\rm step} \ll T_1 , \ T_\phi$.

\begin{table}[htb]
\centering
\caption{\label{decoherence gamma table} Kraus matrix coefficients for decoherence.}
\begin{tabular}{|c|c|c|c|c|}
\hline
$\Gamma_{ma}$  & $a=I$ & $a=X$ & $a=Y$ & $a=Z$ \\
\hline 
$m=1$  & $\frac{1 + \sqrt{1-\gamma-\lambda}}{2}$ & 0 & 0 & $ \frac{1 - \sqrt{1-\gamma-\lambda}}{2}$  \\
\hline 
$m=2$  & 0 & $\frac{\sqrt{\gamma}}{2}$ & $ \frac{i\sqrt{\gamma}}{2}$ & 0  \\
\hline 
$m=3$  & $\frac{\sqrt{\lambda}}{2}$ & 0 & 0 & $- \frac{\sqrt{\lambda}}{2}$ \\
\hline
\end{tabular}
\end{table}

The Pauli channel (\ref{conserving single-qubit pauil channel}) reduces to the depolarization channel ($p_{\scriptscriptstyle X}=p_{\scriptscriptstyle Y} =p_{\scriptscriptstyle Z}$) if
\begin{equation}
\lambda = \gamma (1 - \gamma),
\end{equation}
which occurs when
\begin{equation}
T_\phi = T_\phi^{\rm crit},
\label{symmetric depolarization channel condition}
\end{equation}
where
\begin{equation}
T_\phi^{\rm crit} \equiv \, t_{\rm step}^\frac{\alpha}{1+\alpha} \ (2T_1)^\frac{1}{1 + \alpha}.
\label{symmetric depolarization channel crossover definition}
\end{equation}
Condition (\ref{symmetric depolarization channel condition}) generalizes the usual Markovian-limit  requirement of $T_2 = T_1$ (or $T_\phi = 2 T_1$) and is important because it specifies a crossover in $T_\phi$ beyond which phase-flip errors become subdominant to bit-flip errors (dephasing no longer becomes harmful). 

The decoherence model (\ref{qubit decoherence model}) is extended to $n>1$ qubits by letting the index $m$ in (\ref{CP map}) be the $n$-tuple $(m_1, m_2, \dots, m_n),$ with $m_i \in \lbrace 1,2,3\rbrace$, and
\begin{equation}
E_{(m_1,\dots, m_n)} = E_{m_1} \otimes E_{m_2} \otimes \cdots \otimes E_{m_n},
\end{equation}
where the $E_{m_i}$ are the single-qubit Kraus matrices given in (\ref{Kraus matrices for decoherence}). Then
\begin{eqnarray}
\Lambda\big(\rho\big) &=& \sum_{m_1,\dots,m_n}  E_{m_1} \otimes E_{m_2} \otimes \cdots \otimes E_{m_n} \ \rho \nonumber \\
&\times& E_{m_1}^\dagger \otimes E_{m_2}^\dagger \otimes \cdots \otimes E_{m_n}^\dagger,
\label{multi-qubit decoherence channel}
\end{eqnarray}
where $\rho$ is an $n$-qubit density matrix. Using (\ref{pauli basis kraus expansion}) leads to the twirled multi-qubit decoherence channel
\begin{equation}
{\tilde \Lambda} \big(\rho\big) = \! \! \! \! \sum_{a_1, \dots , a_n}   \! \! \!  p_{a_1 a_2 \cdots a_n} \ a_1 \otimes a_2 \otimes \cdots \otimes a_n \, \rho \,  a_1^\dagger \otimes a_2^\dagger \otimes \cdots \otimes a_n^\dagger, 
\label{multi-qubit decoherence channel}
\end{equation}
where
\begin{equation}
p_{a_1 a_2 \cdots a_n} = \prod_{i=1}^n \, p_{a_i}.
\end{equation}
Here the $p_{a_i}$ are the single-qubit PTA probabilties given in (\ref{PTA decoherence probabilities}).

\subsection{Nonideal CZ gate}
\label{CZ gate section}

Next we consider a two-qubit error channel example, that associated with the  {\sf CZ} gate of Strauch {\it et al.} \cite{StrauchPRL03} for a pair of superconducting qubits or a qubit and resonator bus. The ideal {\sf CZ} gate in the standard basis $\lbrace |00\rangle, |01\rangle, |10\rangle, |11\rangle \rbrace$  is
\begin{equation}
{\rm CZ} \equiv \begin{pmatrix}
1 & 0 & 0 & 0 \\
0 & 1 & 0 & 0 \\
0 & 0 & 1 & 0 \\
0 & 0 & 0 & -1 
\end{pmatrix} .
\label{CZ definition}
\end{equation}
A detailed analysis of the intrinsic errors associated with this gate is given in Ref.~[\onlinecite{GhoshGaliautdinovZhouEtalPRA13}]. Here we consider a particular subset of error processes dominant in the qubit-qubit case (assuming perfect auxiliary $z$ rotations), and parameterize the nonideal {\sf CZ} by the unitary
\begin{equation}
U = 
\begin{pmatrix}
1 & 0 & 0 & 0 \\
0 & \sqrt{1-\mathbb{E}_1}  & \sqrt{\mathbb{E}_1} \, e^{i \phi}  & 0 \\
0 &  -\sqrt{\mathbb{E}_1} \, e^{-i \phi}  & \sqrt{1-\mathbb{E}_1}  & 0 \\
0 & 0 & 0 & - \, e^{i \delta}  
\end{pmatrix} \! ,
\label{nonideal CZ gate}
\end{equation}
where $\mathbb{E}_1 \ll 1$ is the nonadiabatic {\it switching} probability between states $|01\rangle$ and $|10\rangle$, $\phi$ is an arbitrary phase angle, and $\delta \ll 1$ is a controlled-phase error angle. Because $\delta$ is an angle, the leading-order error associated with it is of order $\delta^2 \! .$ The component of (\ref{nonideal CZ gate}) in the $\lbrace |01\rangle , |10\rangle \rbrace$ subspace is assumed to be exactly unitary. The origins of these errors are discussed in Ref.~[\onlinecite{GhoshGaliautdinovZhouEtalPRA13}]. Note that (\ref{nonideal CZ gate}) does not include leakage out of the computational basis. It is useful to relate the size of the errors $\mathbb{E}_1$ and $\delta$ to the state-averaged gate fidelity \cite{ZanardPRA04,PedersenPLA07}
\begin{equation} 
F_{\rm ave}\left(U,U_{\rm target}\right) \equiv \frac{{\rm Tr}(U^{\dagger} U)+
\big| {\rm Tr} \, (U_{\rm target}^{\dagger}U) 
\big|^{2}}{20},
\label{fidelity definition}
\end{equation}
where $U$ is the realized time-evolution operator (\ref{nonideal CZ gate}) and $U_{\rm target}= {\rm CZ}$. We find that the intrinsic gate error 
\begin{equation}
\mathbb{E} \equiv 1-F_{\rm ave} 
\label{gate error definition}
\end{equation}
is, to leading order,
\begin{equation}
\mathbb{E} =  \frac{2}{5} \, \mathbb{E}_1 +\frac{3}{20} \, \delta^2.
\label{leading order gate error}
\end{equation}

To map $U$ to a Pauli channel we write it as
\begin{equation}
U = V \times {\rm CZ},
\end{equation}
where
\begin{equation}
V = 
\begin{pmatrix}
1 & 0 & 0 & 0 \\
0 & \sqrt{1-\mathbb{E}_1}  & \sqrt{\mathbb{E}_1} \, e^{i \phi}  & 0 \\
0 &  -\sqrt{\mathbb{E}_1} \, e^{-i \phi}  & \sqrt{1-\mathbb{E}_1}  & 0 \\
0 & 0 & 0 & \, e^{i \delta}  
\end{pmatrix} 
\label{V matrix}
\end{equation}
is the error. Expanding (\ref{V matrix}) in the two-qubit Pauli basis ${\cal A}_2$ leads to
 \begin{eqnarray}
V &=& \frac{1 +  2 \sqrt{1-\mathbb{E}_1} +  \, e^{i\delta} } {4} \ I
+ \frac{ 1-  e^{i\delta} }{4} \ \big( Z_1 + Z_2 \big) 
\nonumber \\ 
&+&  \frac{ i \sqrt{\mathbb{E}_1} \, \sin \phi  }{2} \ \big( XX + YY \big)
- \frac{i  \sqrt{\mathbb{E}_1} \, \cos \phi }{2} \ \big( XY - YX \big)
\nonumber \\ 
&+& \frac{1 -  2 \sqrt{1-\mathbb{E}_1} + e^{i\delta} }{4} \ ZZ . 
\end{eqnarray}
In the PTA we therefore model the {\sf CZ} gate (\ref{nonideal CZ gate}) by the ideal gate (\ref{CZ definition}) followed by the application of the two-qubit Pauli error channel
\begin{eqnarray}
\rho &\rightarrow& p_{\scriptscriptstyle I}  \, \rho + p_{{\scriptscriptstyle Z}_1} Z_1 \rho Z_1 + p_{{\scriptscriptstyle Z}_2} Z_2 \rho Z_2 + p_{\scriptscriptstyle XX} \, XX \rho XX 
\nonumber \\ 
&+&  p_{\scriptscriptstyle YY} \, YY \rho YY 
+  p_{\scriptscriptstyle XY} \, XY \rho XY + p_{\scriptscriptstyle YX} \, YX \rho YX 
\nonumber \\ 
&+& p_{\scriptscriptstyle ZZ} \, ZZ \rho ZZ ,
\label{CZ error channel}
\end{eqnarray}
where 
\begin{eqnarray}
p_{\scriptscriptstyle I} &=& \bigg| \frac{1 +  2 \sqrt{1-\mathbb{E}_1} +  e^{i\delta} } {4} \bigg|^2 \approx 1 - \frac{ \mathbb{E}_1}{2} - \frac{3}{16} \delta^2 , \nonumber \\
p_{{\scriptscriptstyle Z}_1} &=& p_{{\scriptscriptstyle Z}_2} =  \bigg| \frac{ 1 - e^{i\delta} }{4}  \bigg|^2 \approx \frac{\delta^2}{16}, \nonumber \\
p_{\scriptscriptstyle XX} &=& p_{\scriptscriptstyle YY} = \bigg|  \frac{  \sqrt{\mathbb{E}_1} \, \sin\phi }{2} \bigg|^2 = \frac{\sin^2 \phi}{4} \, \mathbb{E}_1 \le \frac{\mathbb{E}_1}{4} , \nonumber \\
p_{\scriptscriptstyle XY} &=& p_{\scriptscriptstyle YX} = \bigg|  \frac{ \sqrt{\mathbb{E}_1} \cos \phi} {2}  \bigg|^2 = \frac{\cos^2 \phi}{4} \, \mathbb{E}_1 \le \frac{\mathbb{E}_1}{4} , \nonumber \\
p_{\scriptscriptstyle ZZ} &=& \bigg| \frac{1 -  2 \sqrt{1-\mathbb{E}_1}  + e^{i\delta} }{4}  \bigg|^2 \approx \frac{\delta^2}{16} .
\label{PTA nonideal CZ probabilities}
\end{eqnarray}

\section{Bell state preservation}
\label{Bell state section}

\begin{figure}
\includegraphics[width=4.0cm]{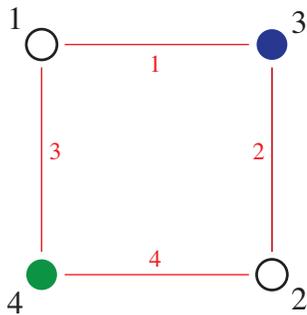} 
\caption{(Color online) Physical layout of the four-qubit system, emphasizing its relation to the surface code. Open circles 1 and 2 are data qubits, the blue (dark gray) filled circle 3 is a $z$-type ancilla (syndrome) qubit, and the green (light gray) filled circle 4 is an $x$-type ancilla. The numbered lines indicate the presence of {\sf CZ} gates and their sequence during each error-detection cycle.}
\label{layout figure}
\end{figure}

In this work we do not simulate an actual encoded qubit, but rather 
a system having a similar (but smaller) stabilizer-measurement circuit. The physical layout of the system we study is shown in Fig.~\ref{layout figure}. This arrangement shows that the four-qubit system can be regarded as a small section of surface code \cite{BravyiArxiv9811052,RaussendorfPRL07,FowlerPRA12}. However, because the number of data qubits is equal to the number of measured stabilizer generators, no logical qubits are encoded.

\subsection{Error correction protocol}

Two {\it data} qubits, 1 and 2, are prepared in a Bell state
\begin{equation}
|B_1\rangle \equiv \frac{|00\rangle + |11\rangle}{\sqrt{2}}
\label{initial Bell state}
\end{equation}
and subjected to amplitude and phase damping. We try to preserve this state by repeatedly measuring the two-qubit stabilizer operator $ZZ$, using ancilla (syndrome) qubit 3, and the operator $XX$ using a second ancilla qubit 4. Ideal simultaneous measurement of these commuting operators projects the data qubits into one of the four Bell states shown in Table \ref{Bell state table}.

\begin{table}[htb]
\centering
\caption{Simultaneous eigenfunctions of the stabilizer generators $XX$ and $ZZ$ with the corresponding
ancilla qubit readouts $x_3$ and $x_4$.}
\begin{tabular}{|c|cc|cc|}
\hline
stabilizer state &  $ZZ$ eigenvalue & $XX$ eigenvalue & $x_3$ &  $x_4$ \\
\hline
$B_1 = 2^{-\frac{1}{2}} \big( |00\rangle + |11\rangle \big)$ &  1  & 1 & 0 & 0 \\
$B_2 = 2^{-\frac{1}{2}} \big( |00\rangle - |11\rangle \big)$ &  1   & -1 & 0 & 1 \\
$B_3 = 2^{-\frac{1}{2}} \big( |01\rangle + |10\rangle \big)$ &  -1 & 1 & 1 & 0 \\
$B_4 = 2^{-\frac{1}{2}} \big( |01\rangle - |10\rangle \big)$ &  -1  & -1 & 1 & 1 \\
\hline
\end{tabular}
\label{Bell state table}
\end{table}

\begin{figure}
\includegraphics[width=8.5cm]{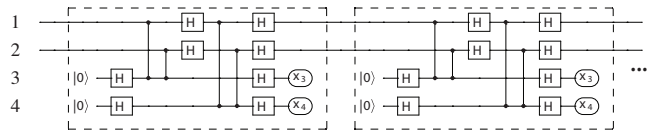} 
\caption{Quantum circuit for the Bell state preservation protocol. Data qubits 1 and 2 are initialized to state (\ref{initial Bell state}). The stabilizer measurement cycle (dashed box) is repeated until the ancilla qubits 3 and 4 return identical values $(x_3,x_4)$ upon measurement for three consecutive cycles. The resulting observed values $(x_3,x_4)$ are used to predict the state of the data qubits, which are then measured in the Bell basis.}
\label{circuit figure}
\end{figure}

The Bell state preservation protocol we simulate is shown in Fig.~\ref{circuit figure}. After preparing the data qubits in the state (\ref{initial Bell state}), a series of error-detection cycles are performed. Each cycle consists of preparing the ancilla qubits in the state $|0\rangle$, and performing an ancilla-assisted measurement of $ZZ$ and $XX$ with the gates shown. Here {\sf H} is the Hadamard gate and the vertical lines are {\sf CZ} gates. We have rewritten the measurement circuit in terms of the {\sf CZ} gate (instead of the {\sf CNOT} gate) for convenience of application to superconducting qubits \cite{GhoshGaliautdinovZhouEtalPRA13}. The ancilla qubits are then measured in the diagonal basis, and the results $(x_3,x_4)$ recorded. The cycle is repeated, without measuring or resetting the data qubits.

In the absence of any errors, $(x_3,x_4)$ will be equal to $(0,0)$ cycle after cycle. However decoherence and gate errors lead to errors on the data qubits, ancilla qubits, or both. In the low-error-rate limit, data qubit errors are identified by a change of
syndrome values from $(0,0)$ to $(0,1), (1,0),$ or $(1,1)$. The new values of $(x_3,x_4)$ will be observed until another error occurs. In this situation we can use Table \ref{Bell state table} to predict the new state of the data qubits. Although we cannot uniquely identify the physical error leading to an observed syndrome change (for example, single-qubit errors $Z_1$ and $Z_2$ both change $|B_1\rangle$ to $|B_2\rangle$), it is still possible to correct the error---apply an operation to return to $|B_1\rangle$---if desired.

Errors can also occur on the ancilla qubits. For example, if the data qubits are in the state $|B_1\rangle$ but a bit-flip error $X_4$ occurs on qubit 4 immediately before readout, then syndrome values $(0,1)$ will be observed, but it will be incorrect to conclude that the data qubits are in state $|B_2\rangle$. However, unless the errors are correlated in time, the syndrome values will return to $(0,0)$ the next cycle. Thus, to protect against ancilla qubit errors it is necessary to run several measurement cycles and ignore single-cycle syndrome changes.

These observations motivate the following Bell state preservation protocol: In each {\it trial}, which represents a single experimental run, the data qubits are initially prepared in state (\ref{initial Bell state}), and the error-detection cycle is repeated until three consecutive ancilla measurements yield the same syndrome values
\begin{equation}
 (x_3,x_4)_{\rm final}. 
\label{stable values}
\end{equation}
Note that (\ref{stable values}) is not necessarily equal to $(0,0)$ because data qubit errors may have occured before the stable values are reached. At this stage one could use (\ref{stable values}) to predict the state of the data and hence the operation required to restore (\ref{initial Bell state})---thereby ``preserving" that initial Bell state---but instead we simply compare that prediction with a measurement of the data qubits in the Bell basis. This measurement verifies the preservation and completes the trail. It occurs after the final syndrome measurement and is the only time the data qubits are measured during the trial. The success probability for this trial is given by the probability $p_B$ of observing the predicted Bell state $|B\rangle$ upon measurement. 

In what follows, we will be interested in the error-correction {\it failure} probability
\begin{equation}
P \equiv {\overline{\ 1-p_B}}
\label{failure probability}
\end{equation}
averaged over many trials, in the presence of decoherence and intrinsic gate errors. The failure probability (\ref{failure probability}) is similar to a logical error rate, and we calculate $P$ exactly---by direct simulation---and by using the PTA.

\subsection{Simulation results}

There are three natural approaches to actually carrying out the PTA simulations: The first is to generate the Pauli errors stochastically and track their effects using stabilizer group techniques \cite{GottesmanArxiv9705052}; this is the standard classically efficient approach used to calculate logical error rates and fault-tolerant thresholds. The second is to generate the Pauli errors stochastically, but track their effects through a full density matrix simulation; this (inefficient) option is simpler to implement but still requires a careful analysis of sampling errors. In particular, it is important to make sure that the error bars due to sampling are smaller than any changes of or differences in $P$ of interest. The third approach, which we follow here, is to fold the PTA-derived Pauli channels back into Kraus maps, which are then implemented exactly. This is also inefficient but avoids the introduction of Monte Carlo sampling errors.

\begin{figure}
\includegraphics[width=10cm]{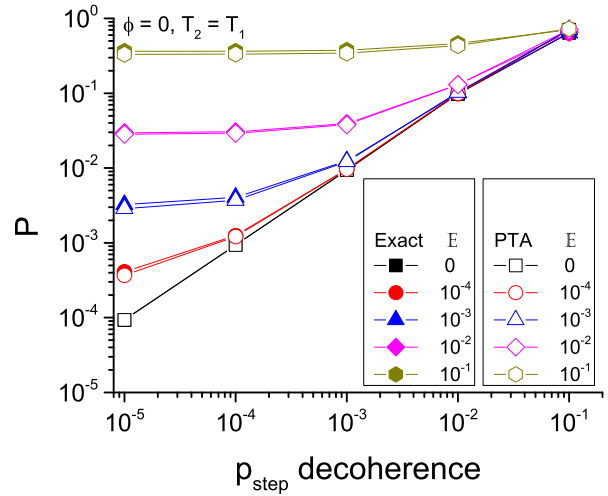} 
\caption{(color online) Error-correction failure probability (\ref{failure probability}) versus per-step decoherence rate, assuming $T_2 \! = \! T_1$. Here $p_{\rm step}$ is the total probability (\ref{pstep definition}) of a decoherence-induced error per $25 \, {\rm ns}$ step of the circuit of Fig.~\ref{circuit figure}. The case where there is decoherence present but no gate errors is shown in black; the agreement is essentially perfect. In addition, $\mathbb{E}$ is the total intrinsic gate error (\ref{gate error definition}) equally distributed over the two error mechanisms (characterized by $\mathbb{E}_1$ and $\delta$) present in the nonideal {\sf CZ} gate (\ref{nonideal CZ gate}).  The explicit values of $\mathbb{E}_1$ and $\delta$ used are provided in (\ref{E1 and delta vs Egate}). The PTA values (open symbols) are hardly distinguishable from the exact results (filled symbols).}
\label{error rate figure}
\end{figure}

\begin{figure}
\includegraphics[width=10.0cm]{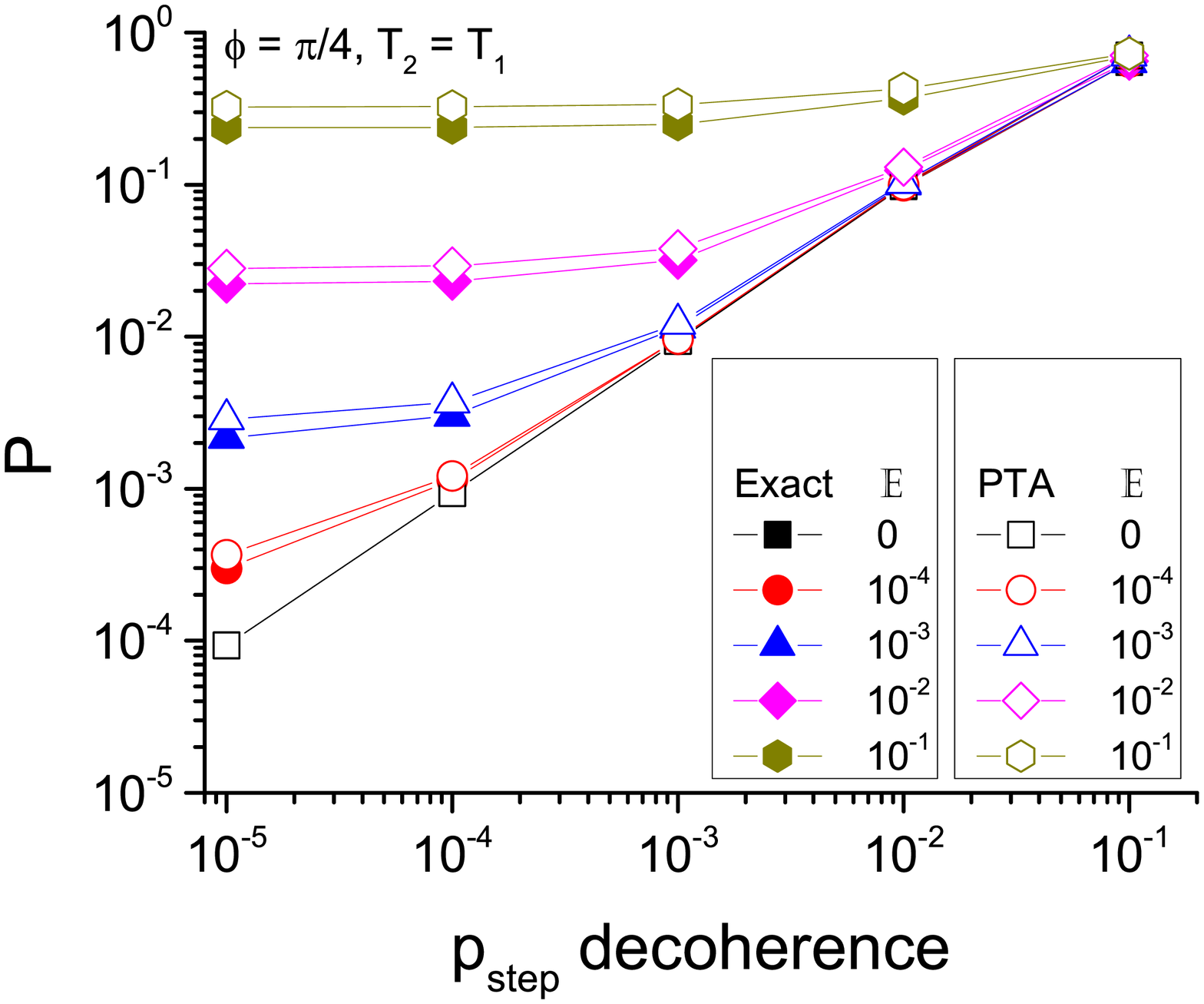} 
\caption{(color online) Same as Fig.~\ref{error rate figure} but with $\phi=\frac{\pi}{4}.$}
\label{P with phi=pi/4 figure}
\end{figure}

\begin{figure}
\includegraphics[width=10.0cm]{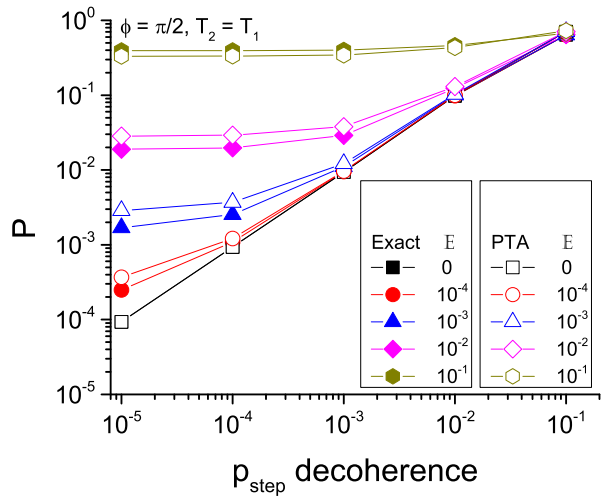} 
\caption{(color online) Same as Fig.~\ref{error rate figure} but with $\phi=\frac{\pi}{2}.$}
\label{P with phi=pi/2 figure}
\end{figure}

We have performed a variety of simulations and find that the PTA is remarkably accurate. Representative results are shown in Fig.~\ref{error rate figure}: Here the error-correction failure rate (\ref{failure probability}) is plotted as a function of the decoherence rate, for a wide range of intrinsic gate errors
(\ref{gate error definition}). The decoherence rate in Fig.~\ref{error rate figure} is changed by varying $T_1$ with $T_2 = T_1$ (and  noise-spectrum parameter $\alpha=0$), and is plotted, not as a function of $T_1$ itself, but as a function of the total probability 
\begin{equation}
p_{\rm step} \equiv p_{\scriptscriptstyle X} + p_{\scriptscriptstyle Y} + p_{\scriptscriptstyle Z}
\label{pstep definition}
\end{equation}
of a decoherence-induced error per step, where the $p_{a}$ are the single-step Pauli error probabilities (\ref{PTA decoherence probabilities}). Note that in Fig.~\ref{error rate figure} (and Figs.~\ref{P with phi=pi/4 figure} and \ref{P with phi=pi/2 figure}, which also have $T_2 \! = \ T_1$), 
\begin{equation}
p_{\rm step} \approx \frac{3 t_{\rm step}}{4 T_1}.
\label{pstep isotropic limit}
\end{equation}
The single-step operation time is  taken to be
\begin{equation}
t_{\rm step} = 25 \, {\rm ns}.
\label{step}
\end{equation} 
Because there are nine steps per cycle in the measurement circuit of Fig.~\ref{circuit figure}, this value of $t_{\rm step}$ results in an error-correction cycle time of $225 \, {\rm ns}$, a value appropriate for superconducting architectures \cite{FowlerPRA12}. 

The nonideal {\sf CZ} gate (\ref{nonideal CZ gate}) is characterized by the three parameters $\mathbb{E}_1,$ $\delta,$ and $\phi$. As discussed in Sec.~\ref{CZ gate section}, $\mathbb{E}_1$ is the nonadiabatic transition probability between the $|01\rangle$ and $|10\rangle$ states, $\phi$  is the associated phase angle, and $\delta$ is a controlled-phase error angle. In Fig.~\ref{error rate figure} we assume that $\phi \! = \! 0$. We do find a small degree of sensitivity to this phase: For example, changing $\phi$ to $\frac{\pi}{4}$ (or $\frac{\pi}{2}$) leads to the results shown in Fig.~\ref{P with phi=pi/4 figure} (or Fig.~\ref{P with phi=pi/2 figure}); in these cases the PTA accuracy is slightly worse. Note from (\ref{PTA nonideal CZ probabilities}) that the value of $\phi$ determines the strength of two-qubit Pauli errors of the form $XX,$ $XY,$ $YX,$ and $YY,$ and that in Fig.~\ref{P with phi=pi/4 figure}, the PTA always overestimates $P$ but in Fig.~\ref{P with phi=pi/2 figure} it underestimates the case $\mathbb{E}_1 = 10\%$ but overestimates the others. The values of $\mathbb{E}_1$ and $\delta$ simulated are such that they each lead to one half of the total intrinsic gate error (\ref{gate error definition}). Therefore we use 
\begin{equation}
\mathbb{E}_1 = \frac{5 \, \mathbb{E}}{4} \ \ \ {\rm and} \ \ \
\delta = \sqrt{\frac{10 \, \mathbb{E}}{3}},
\label{E1 and delta vs Egate}
\end{equation}
which follows from (\ref{leading order gate error}).

\begin{figure}
\includegraphics[width=10.0cm]{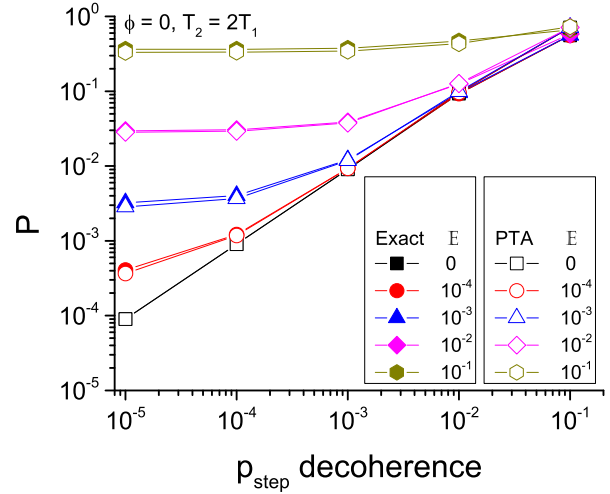} 
\caption{(color online) Error-correction failure probability with $T_2 \! = \! 2 \, T_1$. The values of $\mathbb{E}_1$ and $\delta$ are determined from Eq.~(\ref{E1 and delta vs Egate}).}
\label{P with T1=2 T2 figure}
\end{figure}

\begin{figure}
\includegraphics[width=10.0cm]{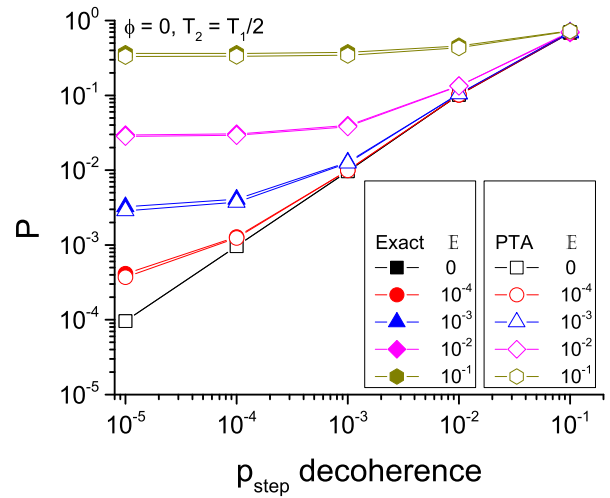} 
\caption{(color online) Error-correction failure probability with $T_2 \! = \! \frac{1}{2} \, T_1$. The values of $\mathbb{E}_1$ and $\delta$ are determined by (\ref{E1 and delta vs Egate}).}
\label{P with T1=0.5 T2 figure}
\end{figure}

In Figs.~\ref{error rate figure} and \ref{P with phi=pi/4 figure} we assume $T_2 = T_1$. This means that the PTA maps the decoherence process to the depolarizing channel with $p_{\scriptscriptstyle X} =  p_{\scriptscriptstyle Y} =p_{\scriptscriptstyle Z}$. In Fig.~\ref{P with T1=2 T2 figure} we show results as in Fig.~\ref{error rate figure} but with $T_2 \! = \! 2 \, T_1$, and in Fig.~\ref{P with T1=0.5 T2 figure} we show results for $T_2 \! = \! \frac{1}{2} \, T_1$. Both show excellent performance of the PTA.

\section{Conclusions and possible directions for future work}
\label{conclusion section}

Although we did not calculate the actual logical error rate associated with an encoded qubit, our results suggest that the performance of the PTA will be similar in that case for {\it small} topological codes. In fact the performance found here is sufficiently good that there is probably little room for further improvement in small codes. However, it must be emphasized that the accuracy of the PTA may not extend to large-distance codes, and this remains an important open question. It would be interesting to go beyond the full Hilbert space simulations reported here and develop analytical or numerical approximation methods, based on more conventional many-body theory techniques, to calculate the logical error rates in large-distance codes for comparison with the PTA and other efficient error models.

It would also be useful to understand {\it why} the PTA works so well. For example, it is known that the ``off-diagonal" terms dropped in the PTA are upper-bounded by the diagonal PTA probabilities $p_{\scriptscriptstyle A}$ \cite{LopezPRA10}. Perhaps this bound on their magnitude, combined with the fact that these terms also come with phase factors that allow for partial cancellation, can explain our observation.

The PTA does not necessarily upper-bound the actual error. It does in some examples but does not in others. However the violation is negligible in the small system studied here, and in our opinion the accuracy of the approximation is more important than the bounding property. In addition, we have shown that the PTA can be simply modified to make it upper-bound the error-correction failure probability $P$: The modification is to take the largest PTA-derived probability 
\begin{equation}
p_0 \equiv \max_A \, \lbrace p_{\scriptscriptstyle A} \rbrace
\end{equation}
for the process in question and construct a new Pauli channel where all (non-trivial) errors occur with probability $p_0$. But the accuracy of the resulting approximation is somewhat compromised.

The combined simplicity and accuracy of the PTA suggests that, at a minimum, it be used as a starting point for further refinements. For example, it might be possible to develop an efficient protocol to {\it perturbatively} include the leading order corrections to the PTA, {i.e.}, the leading-order effects of the neglected off-diagonal terms. Such an approach might also allow one to assess the range of validity of the PTA and whether is breaks down in large codes. 

\begin{acknowledgments}

This research was funded by the US Office of the Director of National Intelligence (ODNI), Intelligence Advanced Research Projects Activity (IARPA), through the US Army Research Office grant No.~W911NF-10-1-0334. All statements of fact, opinion or conclusions contained herein are those of the authors and should not be construed as representing the official views or policies of IARPA, the ODNI, or the US Government.

It is a pleasure to thank Austin Fowler, Joydip Ghosh, and John Martinis for useful discussions. Part of this work was carried out while M.G.~was a Lady Davis Visiting Professor in the Racah Institute of Physics at Hebrew University, Jerusalem.
 
\end{acknowledgments}

\bibliography{/Users/mgeller/bibliographies/algorithms,/Users/mgeller/bibliographies/dwave,/Users/mgeller/bibliographies/control,/Users/mgeller/bibliographies/error_correction,/Users/mgeller/bibliographies/general,/Users/mgeller/bibliographies/group,/Users/mgeller/bibliographies/ions,/Users/mgeller/bibliographies/math,/Users/mgeller/bibliographies/nmr,/Users/mgeller/bibliographies/optics,/Users/mgeller/bibliographies/simulation,/Users/mgeller/bibliographies/superconductors,/Users/mgeller/bibliographies/surface_code,endnotes}

\end{document}